# UpDown: Programmable fine-grained Events for Scalable Performance on Irregular Applications

Andronicus Rajasukumar*, Jiya Su*, Yuqing (Ivy) Wang*, Tianshuo Su*, Marziyeh Nourian*,
Jose M Monsalve Diaz†, Tianchi Zhang*, Jianru Ding*, Wenyi Wang*, Ziyi Zhang*, Moubarak Jeje‡,
Henry Hoffmann*, Yanjing Li*, Andrew A. Chien*†
*University of Chicago, †Argonne National Laboratory, ‡Tactical Computing Laboratories
{andronicus, jiya, yqwang, tssu, marziyehnourian}@uchicago.edu,
josemonsalve2@gmail.com,{tonyztc, jrding, wenyiw, ziyizhang}@uchicago.edu, mjeje@tactcomplabs.com,
{hankhoffmann, yanjingl, aachien}@uchicago.edu

*Abstract*—Applications with irregular data structures, data-dependent control flows, and fine-grained data transfers (e.g., real-world graph computations) perform poorly on cache-based systems. We propose the UpDown accelerator that supports fine-grained execution with novel architecture mechanisms – lightweight threading, event-driven scheduling, efficient ultra-short threads, and split-transaction DRAM access with software-controlled synchronization. These hardware primitives support *software programmable events*, enabling high performance on diverse data structures and algorithms. UpDown also supports *scalable performance*; hardware replication enables programs to scale up performance. Evaluation results show UpDown's flexibility and scalability enable it to outperform CPUs on graph mining and analytics computations by up to 116-195x geomean speedup and more than 4x speedup over prior accelerators. We show that UpDown generates high memory parallelism ($\sim$ 4.6x over CPU) required for memory intensive graph computations. We present measurements that attribute the performance of UpDown (23x architectural advantage) to its individual architectural mechanisms. Finally, we also analyze the area and power cost of UpDown's mechanisms for software programmability.

## I. Introduction

The last several decades have witnessed a rise of "internet-scale" and other big data, producing a growing need for computing systems that process large-scale data efficiently. Graphs are a popular tool in such analysis of irregular data, used to represent complex relationships efficiently in a variety of applications including social networks, world wide web, recommendation systems, bio-informatics and more [30], [34], [40], [51]. Graph computations can exhibit high data-parallelism – exploited in varied software frameworks [20], [42], [61], [63], [69]. However parallelism alone does not translate to scalable, high performance for three key reasons.

First, real-world graphs exhibit highly-skewed structure, so graph computations exhibit unpredictable, indirect memory references. The result is poor cache performance (high miss-rate, over-fetching) [6], [67]. Second, CPU cores with deep memory hierarchies are unable to generate high memory parallelism. The advent of HBM and other AI memories enables DRAM systems to service many references in parallel, delivering high bandwidth. Multi-level caches, tuned for high hit rates, struggle to utilize available memory bandwidth. Finally, CPU cores cannot exploit fine-grained vertex- and edge-level parallelism efficiently [46] and thus, aggregate it into coarse grains. Such aggregation of computation and data movement complicates programming and makes load balancing difficult (e.g., degree variation). Coarse-grained parallelism also limits scalability, as it reduces the available parallelism.

Many graph accelerators have been proposed [6], [11], [14], [29], [55], [64], [77]. Each accelerates a limited set of graph algorithms and often their mechanisms do not generalize across diverse algorithms – even for the same problem, much less to the full space of graph computations. For example, many employ hard-wired intersection units [6], [11], [64], or, they implement hard-wired graph walkers [11], [14], [77], freezing data layout and graph data structures. This approach cannot support the algorithm and data-structure innovation that is acknowledged as the key to continued performance improvement in a post-Dennard, post-Moore era [36]. Many accelerators hard-wire the parallelism model (e.g., vertex-parallel or edge-parallel [18], [29], [55]), limiting application parallelism and flexibility.

We address the research question: *Can we design a programmable accelerator for all graph computations that matches the performance of the hard-wired accelerators?* The response is the design of UpDown, a programmable, event-driven, scalable architecture. UpDown is software-programmable, providing flexibility in data structure, parallelism, and memory access. Consequently, UpDown applications can choose the best algorithms and employ sophisticated data structures. UpDown unifies application-custom software events (fetch_neighbor_list, intersect(u, v) etc.) with hardware events (neighborlist_read_return, write-acks) in a single, efficient event-driven framework. The machine can exploit fine-grained vertex and edge parallelism with hardware parallelism (many threads/accelerator and multiple accelerators/node). This flexible compute parallelism enables the creation of high memory parallelism.

We describe the UpDown accelerator and its software interface (ISA). We evaluate UpDown using a canonical set of graph mining and analysis kernels, comparing it to hard-wired graph accelerators and CPUs. We focus on absolute performance, scalability, and breadth of application. We perform a drill-down study to show the performance contributed by each architectural mechanism. Finally, We characterize the area and power cost of UpDown's software programmability.

Specific contributions of the paper include:

- Design of a software programmable UpDown accelerator that enables efficient fine-grained parallelism. Novel mechanisms include **lightweight threading (LWT), event driven scheduling (EDS), efficient ultra-short thread invocations (UST) and explicit split-transaction DRAM access (SoM)**.
- Experiments that show UpDown is faster than graph pattern mining software (116-fold), and graph analytics software (195-fold) on multicore CPUs. Evaluation that also shows UpDown is (4-fold to 35-fold) faster than hardwired accelerators (across workloads).
- Evidence that shows UpDown's mechanisms enable intelligent data movement to exploit application knowledge to achieve 8x lower memory traffic, 12.5% of cache-footprint, and split-transactions for high memory parallelism, ~4.6x that of a CPU, to deliver up to 3.68TB/s bandwidth. Experiments on graph workloads that show UpDown generates > 1,900 outstanding requests.
- Evaluation that shows UpDown achieves scalable performance (up to 31x on 32 accelerators), including effective support of fine-grained parallelism that improves load balance and delivers 21% better speedup.
- Demonstration of UpDown's flexible programming across 6 workloads that produce ~ 197 diverse events that vary in length from 4 to 1,400 instructions and vary from 0 to 200 DRAM accesses.
- Experiments that decompose UpDown's 23x architectural advantage (vs. simple RISC cores), attributing it to individual architecture mechanisms: 32.1% (LWT), 38.7% (EDS), 1.8% (UST) and 27.3% (SoM).

In Section II we describe graph analytics and mining workloads. Section III frames the problem and describes key elements of our approach. In Section IV, we present the design of the UpDown architecture. In Section V we describe the methodology used for the evaluation in Section VI. Related work is discussed in Section VII. Finally in Section VIII we summarize results and discuss future research directions.

## II. BACKGROUND

Graph processing is a rich class of algorithms from graph pattern mining to graph analytics and graph learning. We use graph pattern mining and graph analytic workloads as representative irregular workloads.

### A. Graph Pattern Mining

Given an input graph $G(V, E)$, Graph Pattern Mining (GPM) involves finding all sub-graphs that match a given $k-$vertex input pattern. The mining problem needs to search the graph and count or list all possible *unique* sub-graphs that match the pattern. As described in [31] software approaches to this problem are of two types - 1) Pattern-oblivious algorithms that build a $k$-level search-tree starting from a vertex and perform isomorphism checks at the leaf [32], [44], [66], [68] and 2) Pattern-aware algorithms that direct the search using symmetry and matching orders derived from the input pattern [31], [43]. Figure 1 depicts three common patterns. UpDown supports both approaches, but for direct comparisons to prior GPM accelerators and software frameworks we use pattern-aware approaches.

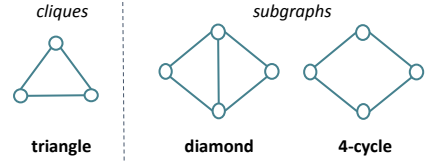

Fig. 1: Common Graph Patterns

### B. Graph Analytics

Graph analytics computation spans decades and includes varied algorithms like search and traversal, community detection, clustering, ranking path detection and so on. Breadth-First Search (BFS) is used to rank graph processing systems in the Graph 500 [16]. PageRank (PR) represents a computation often used as a proxy for graph structure analysis [54]. Jaccard Similarity Coefficients (JS) measures similarity of different graph neighborhoods [33]. We use these three algorithms (BFS, PR and JS) as representative.

These graph workloads have significant parallelism [46] and many software frameworks have been built that exploit it for performance [20], [27], [35], [42], [61], [63], [69]. However, all of these systems have severe limits in either efficiency or scalability.

These limits have produced much interest in hardware accelerators for graph analytics [28].

## III. PROBLEM AND APPROACH

The graph processing problem has facets of data movement, scalability, and programmability. We outline these challenges and our approach to each.

### A. Problem

*1) Inefficient, Low-performance Data Movement:* CPUs employ deep memory hierarchies that exploit data reuse to hide memory latency. Large-scale graph computations often have little reuse; consequently the hardware-managed memory hierarchies fail to hide memory latency, and also fail to generate memory parallelism. We illustrate this with Triangle Counting and Diamond Counting workloads (see Table I).

The poor cache performance is caused by irregular memory references (indirect and random) that produce high miss rates and stymies prefetchers as illustrated in Table I. At these high miss rates, the L1, L2, and L3 caches are ineffective. Further, because they are tuned for low miss rates, the caches typically have a limited number of MSHRs (miss-status handling registers), and thus cannot generate enough requests to utilize high memory bandwidth. The net result is low CPU/core performance as manifested in low instructions per cycle (IPC).

TABLE I: TC and DIA statistics on x86 core [56]

|                           | TC(Yo) | TC(Pa) | DIA(Yo) | DIA(Pa) |
|---------------------------|--------|--------|---------|---------|
| L1D Miss rate             | 17.38% | 19.22% | 33.9%   | 12.99%  |
| L2 Miss rate              | 21.53% | 33.94% | 11.5%   | 17.8%   |
| L3 Miss rate              | 99%    | 99%    | 99%     | 99%     |
| Memory Traffic(GBytes)    | 0.71   | 3.57   | 11.6    | 5.5     |
| Memory Bandwidth(GB/s)    | 0.49   | 0.57   | 0.73    | 0.41    |
| Instructions per Cycle (IPC) | 0.24 | 0.13  | 0.59    | 0.23    |

*2) Poor Scalability:* Graph processing workloads have plentiful fine-grained vertex and edge parallelism [46], but because of architecture limits, existing software frameworks aggregate many vertices per thread or communication for efficiency. Even at this granularity, due to the irregularity in computation and parallelism, these systems struggle with efficiency and scalability [5] (see Figure 2). Demonstrated scaling is poor — only 50% efficiency (Triangle counting) and 25% efficiency (BFS) at 80 threads.

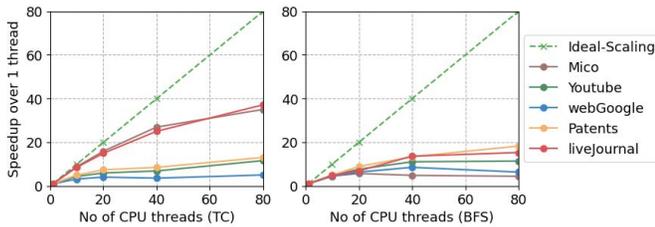

Fig. 2: Speedup vs Threads on Intel Xeon (Skylake) @ 2GHz (dual socket with hyperthreading enabled) on Triangle Counting [12] (left) and BFS [61] (right)

*3) Narrow Acceleration:* Many hardware accelerators for graph analytics and graph mining have been proposed (e.g., [11], [14], [29], [55], [64], [77]). The pattern mining accelerators provide hardware-support to accelerate set operations (e.g., [6], [64]) and depth-first search walkers (e.g., [14]). The analytics accelerators often focus on specific operations (e.g., traversal or vertex update), adding specialized hardware [77]. This approach can deliver performance, but is limited by the specific graph representation (e.g., layout, bit encodings) and parallelism (e.g., vertex, edge).

## B. Approach

*1) Software control for Intelligent Data Movement:* To enable intelligent data movement in UpDown, programs control data movement between local fast memory (scratchpads) and DRAM. This enables application-customized intelligent data movement and efficient use of the fast memory for best application performance. Further, in UpDown, applications can control the size of each DRAM access (no fixed cache block sizes), accessing exactly the data needed.

In addition, to unlock high memory bandwidth, UpDown allows an application to generate an unbounded number of memory requests under program control. When the responses come back, the responses can be customized – software-defined synchronization for the split-transaction interface to DRAM. This memory parallelism is not limited by store buffer or number of MSHRs. These mechanisms enable customized efficient scratchpad use, data movement, and high memory parallelism (UST, EDS, SoM).

*2) Fine-grained parallelism for Scalability:* Graph mining and analytics exhibit irregular parallelism structure and as a result, poor load balance (see Figure 3(left), illustrating skew of 13,700x). These represent challenges for both scalability and efficient hardware utilization. UpDown supports efficient execution of fine-grained thread invocations (10-100 instructions) with direct "register" access to event operands and message instructions. This capability enables UpDown to exploit vertex and edge parallelism – without aggregation to coarse-grain. This enables much higher program parallelism and more uniform work units (thread invocation sizes in Figure 3(right)), supporting better load balance and scalability. This is a stark contrast to threads on server CPU cores – typically scheduled for $> 10$ million instructions. Such short thread invocations require inexpensive thread management, so UpDown provides hardware to schedule/create/destroy threads. The use of fine-grained parallelism allows irregular graph computations to be spread evenly across 2000-fold MIMD hardware parallelism for scalability (EDS, UST, SoM, LWT).

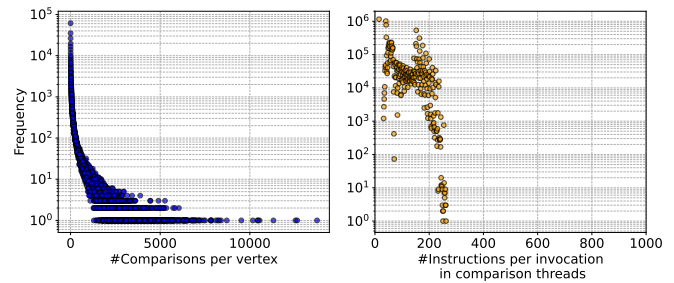

Fig. 3: Graph computing has per-vertex parallelism with size skew of 13.7k-fold (left histogram: comparisons per vertex – TC on Yo dataset). UpDown breaks these irregular chunks into much small thread invocations with lower-skew (20-fold) (right histogram: instructions per invocation – TC on Yo dataset)

*3) Software-defined event computation for broad acceleration – Programmability:* The diversity of graph algorithms and data structure is vast and graph accelerators target specific choices, hardwiring the associated events and computation to achieve high performance.

In contrast, UpDown allows applications to define events flexibly, using software. Combined with efficient, short thread invocations, this enables each application to define the specific events it needs, to exploit a high degree of parallelism and to achieve high performance. Specifically, a program can define, using software, each event's parameters, computation, access to scratchpad, access to DRAM, and when it creates other events. The computation is specified with a general-purpose ISA. These other events can be sent to other lanes or DRAM and synchronize threads. A program can also decide how many instances of the event are created (parallelism). This flexibility enables a broad range of algorithms to be accelerated by

UpDown's general mechanisms, using whatever data structures are most efficient (EDS, SoM).

Together these three dimensions: software-control for intelligent data movement, fine-grained parallelism for scalability and software-defined event computation enable UpDown to achieve extreme graph computing performance. Further, the resulting fine-grained computations scale well to 2000-fold hardware parallelism and support a broad variety of graph algorithms, data structures, and applications. In the next section, we describe the architectural mechanisms that make this performance and generality possible.

## IV. UPDOWN ARCHITECTURE

### A. System Overview

An UpDown node consists of 1 CPU with 32 UpDown accelerators and 8 HBM2e DRAM stacks (Figure 4). The UpDown accelerator sits between the CPU and the DRAM; enabling both UpDown and CPU to access the DRAM. UpDown employs a novel ISA, exposing architectural mechanisms for irregular applications (e.g. graph processing). The ISA supports general-purpose programmability and software performance scalability through efficient fine-grained event processing and parallelism.

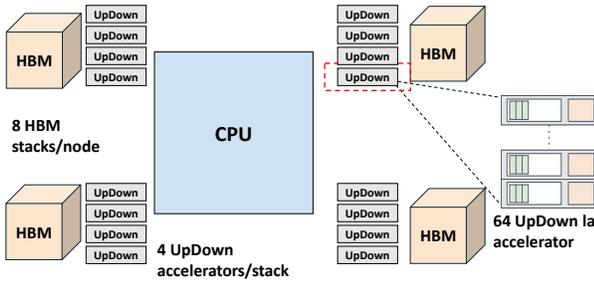

Fig. 4: Each UpDown node has 1 CPU, 32 UpDown accelerators with uniform access to 8 HBM stacks.

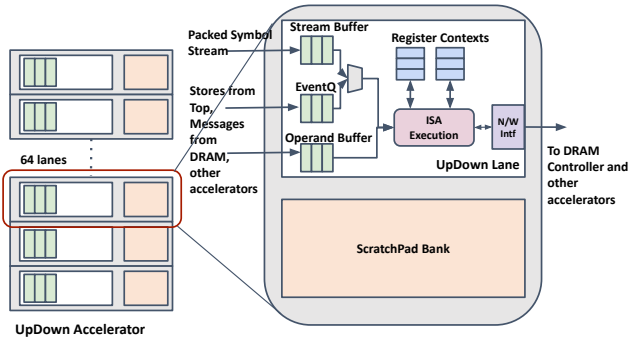

Fig. 5: Each UpDown Accelerator has 64 MIMD lanes with a software controlled scratchpad memory

Each UpDown accelerator has 64 lanes (MIMD) implementing the UpDown ISA [15]. Each lane has a 64KB scratchpad SRAM (see Figure 5). The accelerator is a fraction of the area of a CPU core. Scratchpads have lower latency and energy per access than an equal capacity L1 cache. All 32 UpDown accelerators can access all 8 stacks of DRAM. Software accesses register, scratchpad and DRAM explicitly, enabling flexible control of data movement, memory parallelism, and data locality.

### B. Key Lane Architectural Features

The key ISA [15] and mechanism innovations are realized in each of the UpDown lanes and include:

- **Lightweight Threading (LWT)**: Threads with small register state, fast creation/destruction and high thread parallelism
- **Low-latency Event Driven Scheduling (EDS)**: Hardware thread scheduling and queue management
- **Efficient Ultra-Short thread Invocations (UST)**: Send instructions and message values in register namespace
- **Split-transaction DRAM access with explicit software synchronization (SoM)** Unlimited memory parallelism under software control

*1) Lightweight Threading (LWT):* UpDown supports lightweight thread contexts (16 GPR's, 8 special registers each), unlike CPU threads with large register state, processor state etc. UpDown threads are managed by lightweight mechanisms – 1) creation - by first event (with a special threadID); 2) suspension - using explicit instruction (`yield`); 3) invocation - by events (using the corresponding threadID) and 4) termination - using explicit instruction (`yieldt`) (Figure 6). Given the low cost, each UpDown lane has 128 thread contexts and a local scratchpad memory, accessing it with a load-store interface. Access to DRAM is through program-controlled decoupled messages (see below).

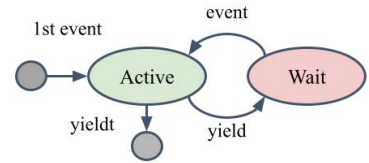

Fig. 6: UpDown Thread Life Cycle. Each time a thread runs (i.e., in active state) is called an invocation.

Both UpDown queues (explained below) and scratchpad memories are mapped into the CPU's address space, enabling it to create events using the event and operand buffers. These events can create threads on the UpDown accelerator. Threads can also be created and invoked by events sent from other UpDown lanes (`send` instructions).

*2) Low-Latency Event Driven Scheduling (EDS):* UpDown employs novel hardware-supported event-driven execution to quickly schedule fine-grained computations. Software-defined events are enqueued (hardware EventQ and Operand Buffer), and trigger single-cycle thread scheduling (see Figure 5). Each event creates a new thread or schedules an existing thread (a thread invocation).

Inexpensive thread creation and invocation enable fine-grain threading and synchronization. UpDown can respond nearly as fast as a hardware finite-state-machine (FSM). This allows programmers to use software threads to respond to hardware events (e.g. dram_read_response, dram_write_ack, etc.). Use of software threads allows the response to be application-defined, the key to acceleration flexibility. For example, see Section IV-D, triangle counting program events such as get_attribute and intersect or examples in Section VI-D.

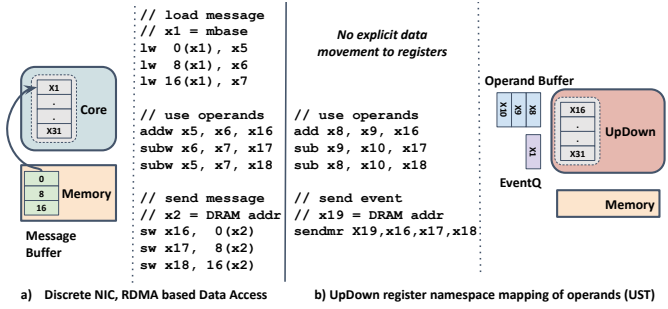

Fig. 7: UpDown maps event operands into register namespace making short computations more efficient

In contrast, conventional cores require 100s to 10,000s of cycles to create or schedule threads [3], [4] due to the cost of trap/interrupt or polling mechanisms. UpDown's approach builds on innovative research machine designs [19] [8], [24], [25], [52], [65].

*3) Efficient Ultra-Short Thread Invocations (UST):* Up-Down mechanisms enable productive work in ultra-short thread invocations ($<10$ instructions). First, UpDown maps message/event values (operands) into the register namespace as in Figure 7. Thread invocations can directly access message data, eliminating instructions to move data from a message buffer into the registers for computation. This design reduces the minimum instructions for a productive thread by up to two-thirds (67%). Many thread invocations with as short as 5 instructions perform useful tasks such as gather, scatter, filter, sum, etc. in high bandwidth flows between DRAM, UpDown lanes and the CPU. Second, UpDown ISA includes instructions (send) to create messages (events) and send them to other lanes or the memory; these actions create computation and memory parallelism with a single instruction.

*4) Split-transaction DRAM access with explicit software synchronization (SoM):* The UpDown ISA provides a novel set of messaging instructions for split-transaction DRAM accesses (sendm). Software accesses DRAM asynchronously with these instructions. A DRAM read/write (1-8 words) is expressed in a single instruction, and the values/ack return in an event from the memory controller. For example, to access data in DRAM, a thread would send a DRAM Read Request event, and then be re-invoked by an event (called continuation) with the desired data. At that point, it can compute directly on the values (mapped into register namespace), or move them to the scratchpad with a single instruction (bcpyol) (Figure 9).

Memory accesses carry a continuation threadID (name of requesting thread), allowing software to handle them flexibly, out-of-order and asynchronous with program state. No expensive hardware look-up structures like Re-order Buffers (ROB) or Miss-Status Handling Registers (MSHR) are required. Up-Down's DRAM access mechanisms can achieve varied data movement tasks efficiently (block-copy, scatter, compact trees, transform matrix etc.). This allows effective software control of data locality (registers, scratchpad, DRAM).

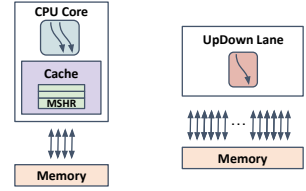

Fig. 8: Traditional cores limit memory parallelism with MSHRs (left), UpDown's Split-transaction DRAM access creates many outstanding memory requests (unlimited) with a single thread (right)

One key advantage of UpDown's is explicit DRAM access-size control. The memory access's size (1-8 words) is specified in the sendm instruction and encoded in the continuation, so when the response comes, the correct number of words are integrated into the register namespace. With control over access size, programs can access DRAM sparsely, avoiding overfetching, and also access multiple words with a single instruction. In contrast, cache-based systems (load-store interface) overfetch because data is moved in cache blocks, and also require load instructions *for each word*.

UpDown's split-transaction memory accesses allow software to create high memory parallelism. Traditional cores generate memory parallelism based on cache misses, limited by MSHR entries (see Figure 8). However in UpDown, a single lane can generate 8-word requests in 3 cycles (loop) for 34 outstanding requests (or 22GB/s DRAM BW); a UpDown accelerator can generate 1000's of outstanding requests (see Figure 9).

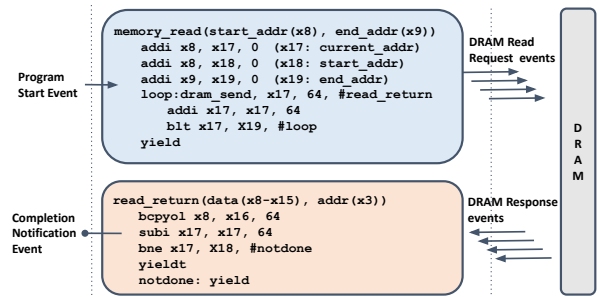

Fig. 9: A 2-event UpDown kernel demonstrating split transaction memory references to DRAM, generating high memory parallelism with asynchronous software synchronization on DRAM responses

### C. Key Mechanisms and Instruction Set Architecture (ISA)

The UpDown accelerator's generality and flexibility come from its novel Instruction-Set Architecture (ISA) and architectural mechanisms that support the features listed in Section IV as described in Table II. Thread control instructions for fast thread suspension, termination (yield, yieldt) and creation (ev, evi, evii) enable low latency hardware thread switching. The event and operands are mapped into the register namespace, and can be used directly by most instructions (arithmetic, logical, load-store, send). This makes short thread invocations efficient. Split-transaction DRAM read/write is done with `sendm`, `sendmr`, `sendmops` instructions. Events can be sent to other lanes with `send`, `sendr`, `sendops` instructions. These send-class instructions format short messages (up to 8 words), providing flexible, variable-length data movement. For scratchpad, in addition to load/store, UpDown adds block-copy (`bcpy`), and streaming compare-n-copy (`cstr`) instructions. Basic synchronization support includes compare-and-swap (`cswp`) instructions on scratchpad memory. The full ISA is available here [15].

TABLE II: UpDown Instruction Set Architecture Highlights

| Category | Execution Mechanisms and Instructions |
|---|---|
| Lightweight Threading | Thread spawn/invoke: Msg arrival event, send msg; Thread yield/destroy: (`yield`, `yieldt`) |
| Event-driven Scheduling | Event and Operand Queues; hardware management w/o any instructions |
| Ultra-Short Threads | Register naming for Msg Operands, use in add-class, send, and load-store (scratchpad memory) operations |
| Split-transaction DRAM | DRAM send instructions (`sendm`, `sendmops`, `sendmr`); DRAM responses come to software-defined event label |
| Messaging (incl DRAM) | Single instruction sends of 1-8 words: `send`, `sendr`, `sendops`, `sendm` (DRAM), `sendmops` (DRAM) |
| **Traditional Instructions** | add-class arith and logic (`add`, `sub`, `and`, `or`) and control flow (`beq`, `ble`, `bgt`) |
| Scratchpad Memory | load-store (`movlr`, `movrl`), copy block (`bcpy`), streaming compare-n-copy (`cstr`) |
| Synch (scratchpad) | compare-and-swap (`cswp`) |

### D. Program Example (Triangle Counting)

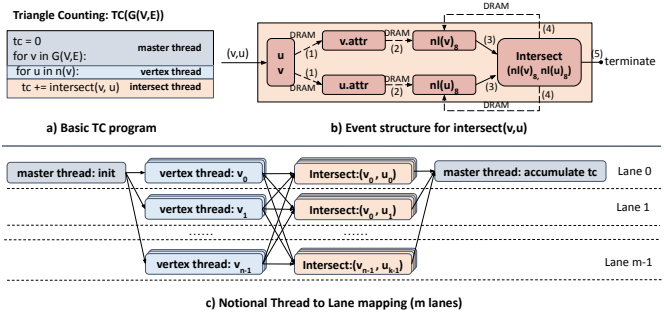

Fig. 10: Triangle Counting on UpDown

We describe the triangle-counting (TC) program to illustrate UpDown's software programmability and expression of fine-grained parallelism. Conceptually all vertex and intersections can be executed in parallel (see Figure 10a). Coarse-grained approaches parallelize groups of vertices (outer loop). UpDown's fine-grained parallelism allows *all vertex iterations* and *all intersection iterations* (inner loop) to be executed in parallel.

Even within the numerous intersection iterations (threads), there is more fine-grained parallelism; we show its event-structure in Figure 10b. Each intersection thread fetches the vertex attributes ($u$ and $v$) from DRAM (1), saving to scratchpad. These attributes (vertex-degree, id and neighbor-list pointer) are used to fetch the neighbor lists $nl(v), nl(u)$ in segments of 8, $nl(v)_8, nl(u)_8$ which are the used to calculate the intersection (3). Depending on the termination condition of the segment, the same event will fetch the next segment for the corresponding neighborlist ($nl(v)_8$ or $nl(u)_8$) (4) until symmetry conditions or end-of-neighborlist conditions are met. Each DRAM access yields the computation resource (lane) to other thread invocations.

From a high-level (Figure 10c), there's a hierarchy of vertex and intersect threads that are spread across the UpDown lanes. The master is launched from the CPU and in turn launches vertex threads. After fetching the neighbor vertices ($u \in n(v)$), the vertex thread launches intersect threads. The intersect threads update the local triangle count ($tc\_val$) atomically in scratchpad before termination. Finally, the master thread performs a reduction to complete the computation. The fine-grained parallelization produces balanced, high lane utilization (see Section VI), enabling good performance scaling.

## V. METHODOLOGY

### A. Workloads and Datasets

For evaluation of the UpDown architecture we use graph mining and graph analytics computations (Table III). These computations are run on a variety of real-world graphs (Table III). For JS, we use smaller datasets ($O(V^2)$ complexity [33]).

TABLE III: Graph Workloads

| Graph Pattern Mining | Graph Analytics |
|---|---|
| Triangle Count (TC), Diamond Count (DIA), 4-cycle (4CYC) | Breadth-First Search (BFS), PageRank (PR), Jaccard Similarity Coefficients (JS) |

TABLE IV: Real-World Graph Datasets

| Dataset | #Vertices | #Edges | #Size | #Max deg | #Avg deg |
|---|---|---|---|---|---|
| MiCo(Mi) [22] | 96K | 1.1M | 19.3MB | 936 | 11 |
| Youtube(Yo) [72] | 1.1M | 3M | 72MB | 28,754 | 5 |
| Web-Google(Wg) [38] | 875K | 5.1M | 87MB | 6,332 | 3 |
| Patents(Pa) [37] | 3.8M | 16.5M | 340MB | 793 | 9 |
| LiveJournal(Lj) [72] | 4M | 34.7M | 622MB | 14,815 | 17 |
| bio-CE-CX(Bcc) [57] | 15.2K | 246K | 5.19MB | 375 | 32 |
| bio-grid-human(Bgh) [57] | 9.4K | 62.4K | 340KB | 616 | 13 |
| Erdos-Renyi-400x16k-p0.01(Er) [23] | 16.4K | 65.8K | 332KB | 14 | 4 |

### B. Simulation Model and Configurations

We use gem5 [7], [41] with standard core model (x86 OOO), cache, and network components (see Table V). We wrote and integrated a cycle-accurate instruction-level simulator for the UpDown accelerator (lanes and scratchpad). We use DRAMsim3 [39] to model the HBM2e stacks.

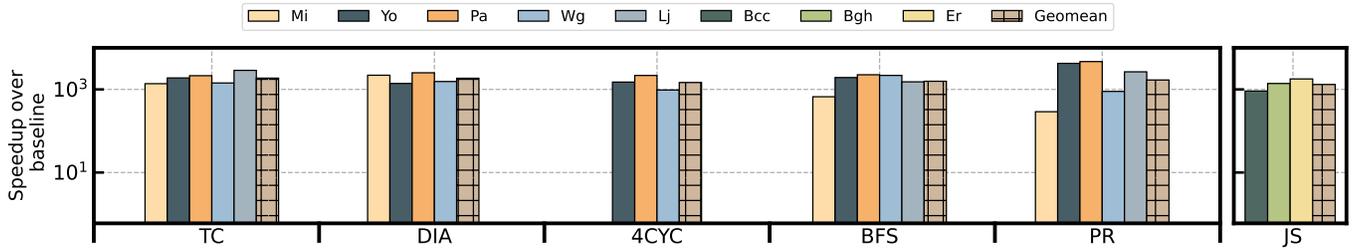

Fig. 11: UpDown speedup. Varied workloads and graphs. Rightmost bar for each cluster is Geomean.

The UpDown system includes an x86 Intel (Skylake) OOO CPU connected to HBM2e stacks @460GB/s through a cache hierarchy (see Table V). The CPU is used to start the computation, writing an event to an UpDown lane. Each UpDown accelerator has 64 UpDown lanes, and an UpDown node has 32 accelerators and 8 HBM2e DRAM stacks. To generate the paper results we performed ∼ 750 simulations that total ∼ 30,000 CPU hours (approximately 3.5 CPU-years).

TABLE V: System Components

| Component | Specification |
|---|---|
| CPU | x86 core Intel (Skylake), OOO core, @2GHz<br>L1: 64KB DCache, 32KB ICache per core,<br>L2: 256KB unified per core<br>L3: 8MB shared |
| UpDown Accelerator | 64 UpDown Lanes + 4MB scratchpad (64KB per lane) @2GHz |
| DRAM Memory | 8x channels HBM2e @460GB/s |

TABLE VI: Experiment Configurations and Metrics

| Configuration | Description |
|---|---|
| UpDown | A node with CPU, 32 UpDown Accelerators and 8x HBM2e DRAM stacks |
| **Deep-dive Configurations** | **Description** |
| PE | In-Order Lane |
| PE + SoM | Add Split-Transaction DRAM Access (SoM) |
| PE + SoM + LWT | Add multiple Lightweight thread contexts (LWT) |
| PE + SoM + LWT + UST | Add efficient Ultra-Short Thread invocations (UST) |
| UpDown Lane | Add Low latency Event-Driven Scheduling (EDS) |
| **Metric** | **Definition** |
| Runtime (seconds) | Execution time |
| Speedup | Relative Performance |
| Lane Utilization | Fraction of Runtime spent in execution |
| Memory Traffic (GB) | Total DRAM read + write bytes |
| Memory Bandwidth (GB/s) | Memory Traffic / Runtime |

### C. Reference Software Frameworks and Graph Accelerators

For uniform comparison across mining and analytics software frameworks and graph engines, we use single-threaded implementations, GraphZero (20-thread) [43], [44] for graph mining (TC, DIA, 4CYC) and

Ligra (20-thread) [61] for analytics (BFS, PR) as baselines. We use open source benchmarks from GraphMiner [12], [13].

We use published speedups for FlexMiner [14], NDMiner [64], FINGERS [11], Graphicionado [29], GraphPulse [55] and DepGraph [77] compared to the software frameworks, adjusting for CPU clock frequency.

JS is compared to SISA [6] (best available, projecting optimistically - linear).

## VI. EVALUATION

We evaluate UpDown, showing that its mechanisms deliver scalable, high performance. With graph workloads (Table III) and real-world graphs (Table IV), we show how the architecture realizes programmable, broad acceleration.

### A. Overall Performance

We compare UpDown to a single x86 thread to provide a uniform basis for performance comparisons. The results show UpDown delivering geomean speedups of 1,461-1,865x across varied workloads (see Figure 11)). Per workload, best speedups are 2,868x (TC), 2,507x (DIA), 2,162x (4CYC) and 2,248x (BFS), 4,665x (PR), and 1,782x (JS).

Compared to thread-parallel software on multicore, UpDown achieves geomean speedup of 116x (GraphZero) and 195x (Ligra), with maximums speedups of 212x and 434x respectively (see Figure 12). When compared to graph accelerators, UpDown demonstrates performance improvements of 4x to 35x across workloads. The ✗ indicates that comparison is not possible due to lack of data/implementation.

Overall, UpDown's absolute performance exceeds software frameworks by 100-fold, and outperforms existing graph accelerators by a significant margin.

### B. Intelligent Data Movement

UpDown addresses irregular data movement challenges by providing software-control of data movement and the ability to generate high memory parallelism. The benefit of UpDown's software control can be seen in a simple matrix access example. When accessing a column in a matrix with row-major layout (e.g., c array), a cache would load a 64-byte block for each 8-byte word. UpDown programs can specify DRAM access size and thus, need to only fetch a single word per column value. This produces the comparison in Figure 13; UpDown avoids overfetching, reducing memory traffic 8-fold. Further, under software control, the column data can be packed in the scratchpad, so compared to a typical cache, the column data requires less than 12.5% footprint (no tag overhead).

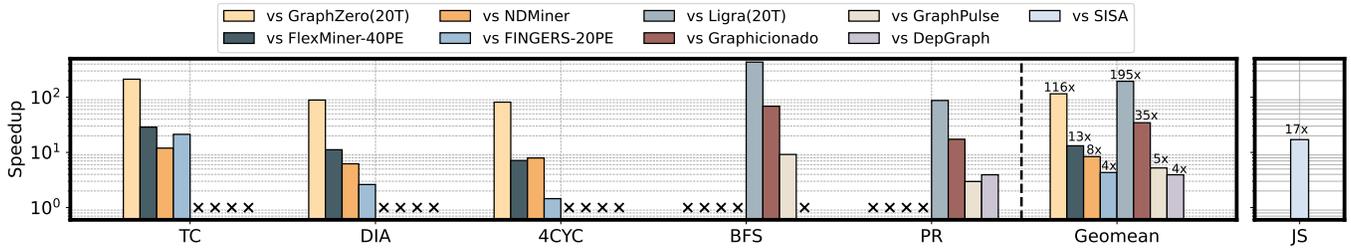

Fig. 12: UpDown performance comparison with software (multicore) and hardwired accelerators. ✗ indicates comparison not available.

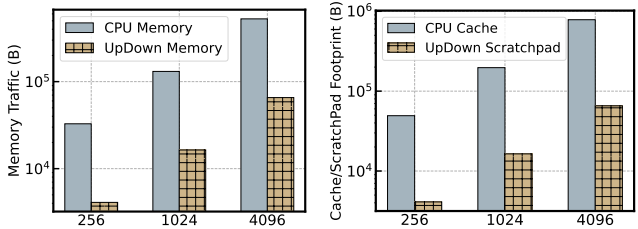

Fig. 13: Memory Traffic and Scratchpad Footprint (CPU, UpDown) for a column access in a row-major dense NxN matrix

UpDown enables a single thread to have many outstanding memory requests, and multiplies that with inexpensive thread parallelism to achieve high memory parallelism. As described in Section IV-B4, each thread can issue an unlimited number of memory requests (100's even 1000's), and perform custom synchronization/processing for each response. Thus the bandwidth achievable by a single thread is limited only by its instruction issue bandwidth and the maximum request size (8 words). This is illustrated in the left-most ramp in Figure 14. Each 8-word DRAM request requires 3 instructions, with 3 instructions for each DRAM response a single thread or lane can achieve 22GB/s.

Next, thread parallelism can be used to scale memory parallelism. UpDown supports efficient thread creation (3 cycles/thread, one lane can create 667 Million threads/sec). Using many threads, Figure 14 shows how memory bandwidth can be scaled to 458GB/s (1 accelerator, 64 lanes), all the way to 3.68 TB/s (32 accelerators).

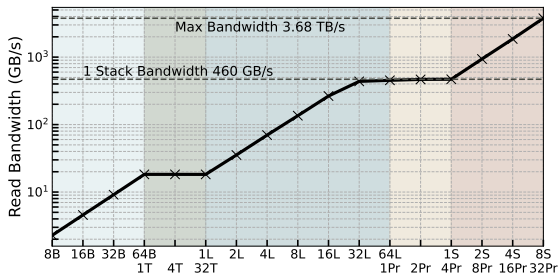

Fig. 14: UpDown DRAM Bandwidth vs (transfer size, threads, lanes, accelerators, stacks)

An alternative way of understanding this is in terms of outstanding memory requests. UpDown's design enables threads to generate a large number of outstanding memory requests that saturate the DRAM stack's memory parallelism. Given typical DRAM latencies, each lane achieves DRAM request parallelism of $\sim 34$, and $\sim 2,200$ for UpDown accelerator. For 32 accelerators, potential request concurrency exceeds 70,000, and is limited by the memory hardware parallelism to $\sim 6,000$-fold on 8 HBM2e stacks. For comparison, we estimate this is $\sim 4.6x$ the memory parallelism achieved by the 56-core Intel Sapphire Rapids that is STREAM benchmarked at 700GB/s with 4 HBM2e stacks (43% of peak) [45].

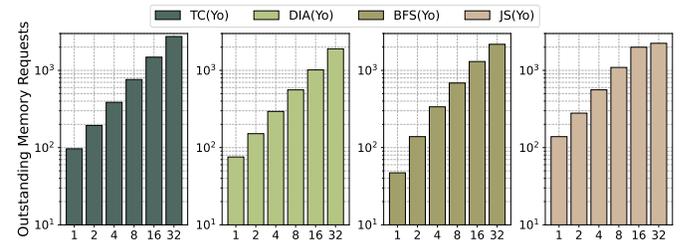

Fig. 15: UpDown Outstanding DRAM Requests (memory parallelism) for 1 - 32 accelerators (64-2048 lanes); typically 34 per lane.

Exploiting UpDown's program-controlled data movement in graph workloads entails customization for data structure and access pattern. TC, DIA, BFS use edgelists, JS uses compress-sparse-rows (three vectors); BFS uses split vertices to manage irregularity, TC streams both neighborlists from DRAM for each intersection, and in contrast, JS software-caches one in scratchpad and streams the other. Remarkably, despite this diversity, all are able to achieve high memory parallelism with UpDown's mechanisms (see Figure 15). The workloads presented average $\sim$ 1,900-2,750 outstanding memory requests, a large fraction of the physical parallelism available of less than 6,000. UpDown's high achieved memory parallelism is the key to its scalable, high performance.

### C. Scalable Performance

The key element of UpDown's approach to performance is scalability based on fine-grained parallelism. We present scaling results for TC, DIA and BFS. UpDown achieves linear speedup for TC and DIA, and near linear for BFS, a more challenging benchmark (see Figure 16). Exploiting fine-grained parallelism enables performance scalability as high as 31x on 32 accelerators (2048 MIMD lanes) for a variety of workloads. We explore how UpDown's fine-grained parallelism support makes this good scaling possible.

UpDown programs use many threads and execute them with ultra-short thread invocations. Table VII presents counts of threads, thread invocations, and invocation sizes. We choose Yo because it's a difficult graph to get good speedups. The statistics show that UpDown programs have numerous threads that execute *very* short invocations (15-68 instructions). See also TC thread-invocation instruction count distribution in Figure 3(right). Breaking computation into short invocations increases parallelism and improves workload balance. For example, in TC, this increases parallelism by exploiting both vertex and intersection level parallelism achieving 21% increase in speedup with 32 UpDown accelerators from 1,482x to 1,885x (Figure 17). Another way to understand this is to examine lane utilization (see Figure 18). Coarser-grained parallelism (thread-per-vertex) produces lane utilization skew of ∼ 51% to 93% (32 accelerators). UpDown's finer-grained parallelism (thread-per-intersection) reduces skew 4-fold to ∼ 80% to 90%. The reduced skew and high utilization enable improved scalability. UpDown's ability to support fine-grained parallelism has benefits for irregular applications generally (not just graphs).

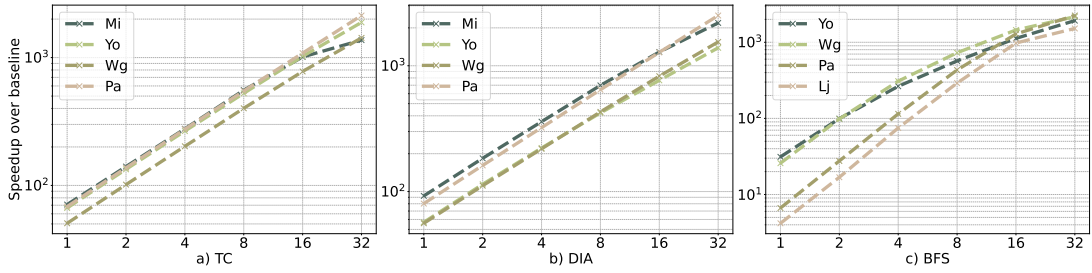

Fig. 16: UpDown Speedup for TC, DIA and BFS. Various datasets. 1-32 accelerators (64 - 2048 lanes)

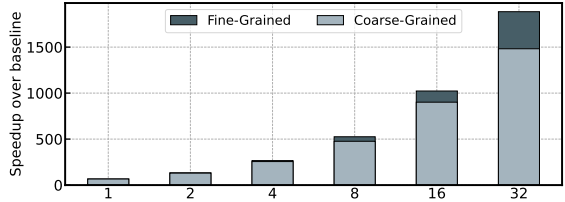

Fig. 17: Fine-grained parallelism improves speedup when scaling 1-32 accelerators on TC(Yo)

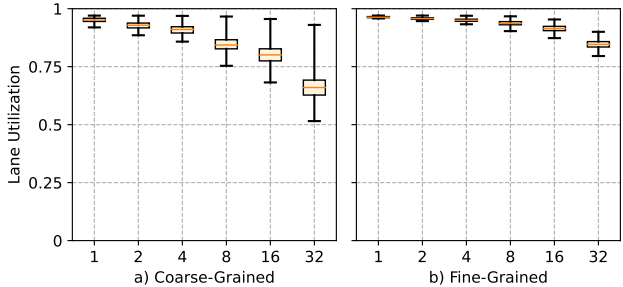

Fig. 18: Fine-grained parallelism improves Lane Utilization by reducing skew. 1-32 accelerators on TC(Yo).

TABLE VII: Thread & Invocation Statistics on Yo (Er for JS)

| Appl. | # of Threads | # of Invocations | Avg Insts / Invocation | Avg MemoryRef / Invocation | Avg DRAM Bytes/ MemoryRef |
|---|---|---|---|---|---|
| TC | 3,978,416 | 51,232,414 | 52.35 | 0.48 | 34.40 |
| DIA | 3,978,416 | 482,155,377 | 67.56 | 0.50 | 59.61 |
| 4CYC | 16,245,100 | 2,318,790,317 | 71.28 | 0.50 | 62.12 |
| PR | 19,102,752 | 27,498,715 | 14.97 | 0.21 | 27.47 |
| BFS | 8,571,634 | 20,059,589 | 17.41 | 0.39 | 39.95 |
| JS | 132,483,761 | 418,166,953 | 28.38 | 0.38 | 62.34 |

### D. Flexible Programmability

A critical approach for UpDown's high performance across varied workloads is the flexible programmability of events, enabling algorithm and data-structure innovation. The 6 workloads consist 197 software-defined events to achieve scalable, high performance. In Figure 19, we illustrate this diversity with selected event structures from JS and PR; these two workloads alone have 26 software-defined events. More are illustrated in the TC description in Section IV-D. In contrast, most graph accelerators implement hardwired operations (e.g., set intersection, set difference, path-prefetching, symmetry breaking, [64], and fine-grained scheduling) for acceleration, limiting their breadth of benefit.

UpDown uses its flexible software interface to define events (e.g., vertex_updates, intersections), and unifies it with hardware events (e.g., dram_reads, dram_read_response). This allows programs to define events that match their data structure, algorithm and irregularity needs. This software customization is key to UpDown's broad applicability and performance benefit.

Another perspective on UpDown's flexibility in events can be seen in diverse workload use. Events vary widely in thread sizes, invocation sizes, and number of DRAM accesses per invocation, both across workloads and even within workload. For example, in Figure 20, the average instructions per invocation varies 3-fold across workloads. However it varies even more within each workload, from 4 instructions to 750 to 1500 instructions. In another dimension, the number of DRAM references per invocation also varies widely, but here the variation is large in only two of the workloads (JS, 0-40 and BFS, 0-200). Such irregular structure in computation and memory references makes it difficult to design a single hardware accelerator, yet UpDown's flexible software-defined events can span the diversity, providing broad acceleration.

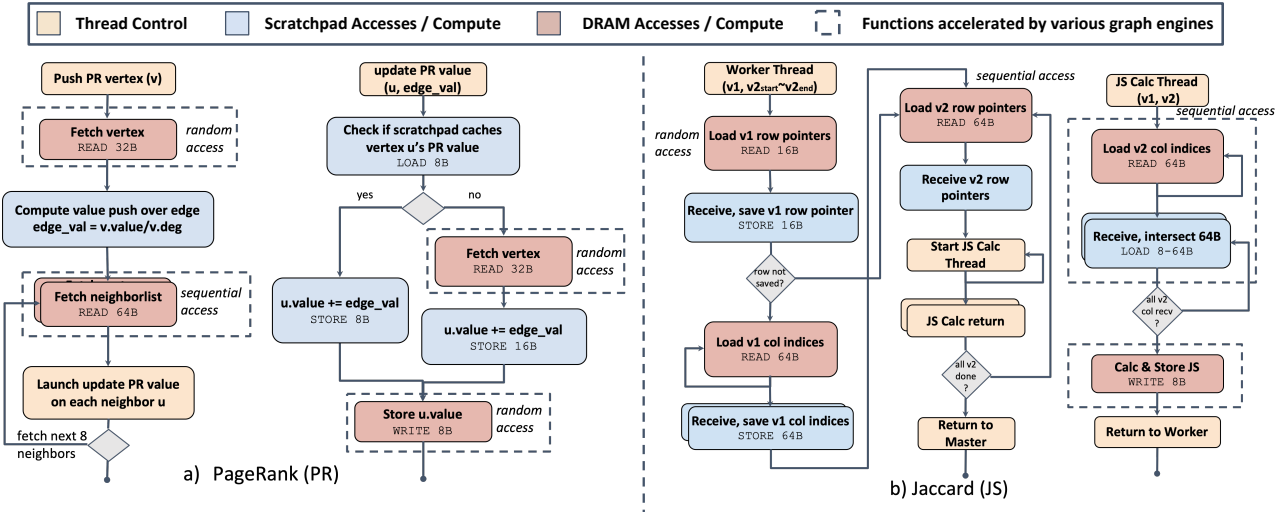

Fig. 19: Software Event Structure of two exemplar programs (JS and PR); illustrates 26 of the 197 diverse events from the workloads

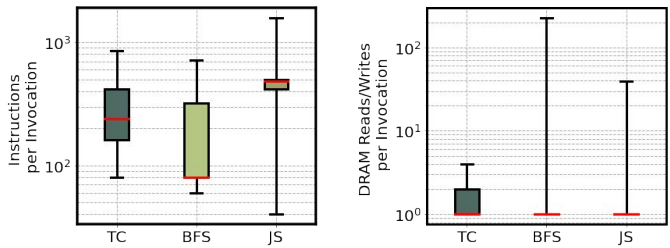

Fig. 20: Distribution of Instructions per invocation (left) and DRAM accesses per invocation (right) (TC, JS and BFS). Medians vary significantly across applications and more widely across events within (10-1000x)

### E. Deep Dive

Our deep dive experiments dissect performance, assigning performance benefit to each UpDown architecture mechanism. Figure 21 shows UpDown's 23x (10-43x) architectural advantage relative to a 64-lane accelerator built from PEs (akin to in-order RISC cores). We divide this overall benefit by measuring application performance on variants of the UpDown accelerator. (configurations in Table VI).

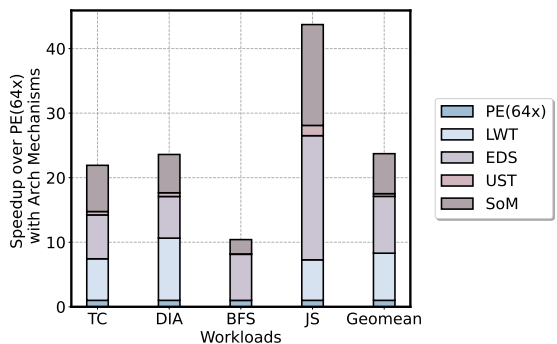

Fig. 21: Speedup Decomposition of UpDown architectural mechanisms on TC, DIA, BFS and JS

First, lightweight threading (LWT), contributes 32.1% (14-41%). As a single thread does not fully utilize the lane, multiple thread contexts increase lane utilization. Second, low-latency event driven scheduling (EDS) eliminates overhead for event scheduling, removing the cost of interrupt handling, thread context switches. It contributes 38.7% (27-68%). Third, ultra-short threads (UST) eliminate message handling overhead, making short invocations efficient, providing a 1.8% (1-3.6%) benefit. As shown in Table VII, some workloads average as few as 14 instructions per invocation. Fourth, split-transaction DRAM accesses with software synchronization (SoM) enables high memory parallelism, contributing 27.3% (21-36%). SoM eliminates the MSHR bottleneck.

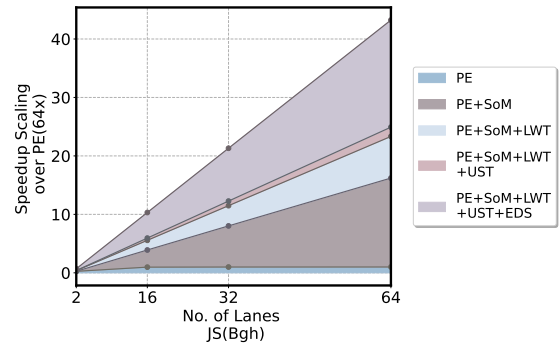

Fig. 22: Speedup from UpDown architectural mechanisms on JS(Bgh), 2 to 64 Lanes

Finally, we summarize UpDown's architectural mechanisms' incremental performance impact in Figure 22 with an example scaling from 2-64 lanes.

In this example, performance contributions from largest to smallest are (45.0%(EDS), 36.6% (SoM), 14.7%(LWT) and 3.7%(ST)), and scale with the number of lanes. The overall benefit at the accelerator-level is 43x, relative to a 64 PE baseline.

### F. Cost: Area and Power

**Area:** We estimate the Area and Power of the UpDown accelerator based on 28nm implementation and synthesis using Synopsys Design-Compiler on 28-nm TSMC process [1].

Critical components include the Datapath, EventQ, Operand-Buffer, Thread registers and 4MB Scratchpad. Applying technology scaling for SRAM [9] and logic [62] to 7nm (see Table VIII) shows that Scratchpad dominates the area.

TABLE VIII: Area and Power

|  | Component | Area[28nm] ($mm^2$) | Area[7nm] ($mm^2$) | Fraction |
|---|---|---|---|---|
| Accel Lane | Dpath,EventQ, OperandB,Regs | 0.1791 | 0.0269 | 0.67% |
| Accelerator | 64 Lanes | 11.46 | 1.72 | 42.89% |
| Scratchpad | 4MB SRAM | 12.56 | 2.29 | 57.11% |
| Accelerator System | 64 lanes + 4MB | 24.03 | 4.02 | 100 % |
| Sapphire Rapids [47] | 1 Core + L1/L2/L3 | n.a | 28.6 | n.a |

|  | Component | Power[28nm] ($mW$) | Power[7nm] ($mW$) | Fraction |
|---|---|---|---|---|
| Accel Lane | Dpath,EventQ, OperandB,Regs | 20.758 | 5.94 | 0.65% |
| Accelerator | 64 Lanes | 1328 | 380 | 41.82% |
| Scratchpad | 4MB SRAM | 1,848 | 528 | 58.18% |
| Accelerator System | 64 lanes + 4MB | 3,176 | 908 | 100% |
| Sapphire Rapids [47] | 1 Core + L1/L2/L3 | n.a | 6,071 | n.a |

Compared to Intel's Sapphire Rapids (4x400$mm^2$ tiles in 7nm), each tile has 14 cores, for 56 cores total and 28.6$mm^2$/core [47]. The UpDown accelerator is 4$mm^2$ or about 1/7 the size of a core. Comparing 32 UpDown accelerators to the Sapphire Rapids processor (32x4=128$mm^2$/1600$mm^2$) shows 32 UpDowns are only 8% of the Sapphire Rapids logic area.

**Power:** We use CACTI [48] for memory and [62] for logic power estimation in 7nm. Power is dominated by SRAM leakage. At system level, 32 UpDown accelerators are 29/350 = < 8.3% of a Sapphire Rapids CPU; A single UpDown accelerator power (64 lanes) is 14.9% of one of Sapphire Rapids' 56 cores.

### G. Cost of General Programmability

UpDown includes no graph-specific or data-structure specific primitives – a stark contrast to other graph accelerators. It provides a variety of fine-grained and parallelism primitives. To put UpDown's area and power in context, we assess the *cost of generality* by comparing the cost per unit performance in Figure 23. Since these accelerators are implemented on varying technology nodes, we use published area and power numbers and speedups from Figure 12 to derive this comparison.

Compared to CPU, UpDown is 356x more area-efficient (0.9$mm^2$/unit perf.) and 686x more power-efficient (204$mW$/unit perf). 32 UpDown accelerators could be added to a CPU with $\sim 8\%$ area/power overhead. Compared to graph accelerators, UpDown is 5x less area-efficient and 2x less power-efficient than NDMiner (best case). Note DepGraph [77] is an outlier on area because it is a graph specific

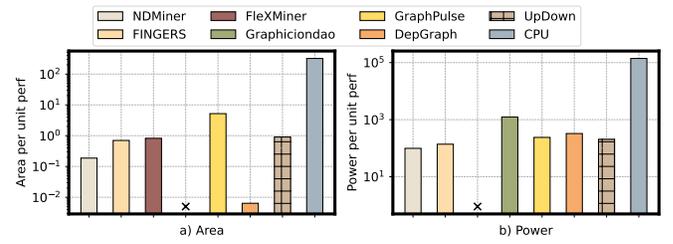

Fig. 23: Cost of General Programmability for Graph Processing: cost/performance from accelerators to CPUs using a) Area and b) Power [14], [64] [55] [29] [77] [17]. ✗ indicates unavailable data

prefetcher, that does minimal computation. While some of these accelerators are highly area and power efficient, their lack of generality implies that an ensemble of accelerators is required to cover the same range of applications.

## VII. RELATED WORK

### A. Hardwired Graph Accelerators

Numerous research projects proposed accelerators for graph mining [6], [11], [14], [32], [64], [70], [73] and graph analytics [18], [29], [55], [77], achieving good speedups. However, they generally hardwire data-structures, algorithms and graph-specific mechanisms (like symmetry breaking, hardware accelerated set operations, accelerated DFS walkers, high-degree vertex caching etc.) limiting their generality. The rapid algorithmic innovation in graph processing suggests the value of flexible acceleration. UpDown takes a different approach, providing software-programmable, flexible acceleration. UpDown mechanisms described in Section IV target the key properties needed, e.g., supporting large numbers of ultra-short threads (compute parallelism) and maximizing outstanding memory requests (memory parallelism) to achieve high performance on irregular computations in general.

### B. General Processing In Memory

Continued pressure on memory performance [49] has produced a resurgence in Processing in Memory (PIM) / Processing near Memory (PNM) architectures. These architectures focus on data movement reduction, typically using simple in-order cores set in proximity to memory [2], [6], [50], [53], [76], [78]. These designs limit the data addressability/access of individual cores to a local row, chip or stack. This makes effective programming for large graph data sets difficult. In UpDown, all accelerators can access all 8 stacks of DRAM.

### C. Optimized Memory Hierarchy Management

Numerous efforts to improve cache hierarchies for irregular applications have been proposed. Custom prefetch accelerators [74], [75], data-structure aware path-prefetching [77] focus on maintaining data supply rate through intelligent prefetching techniques. Other innovative approaches [59], allows cache-events to trigger custom data movements using callback functions. However, these techniques are implemented in hardware and generally operate on cache-blocks, limiting the generality and range of improvement. In contrast, UpDown provides full software control over data movement size and strategy, exploiting application information and focusing on primitives to make this efficient.

*D. Software Graph Processing Frameworks on CPUs/GPUs*

Graph Analytics and Graph Mining Software Systems have been proposed for large-scale graph processing on CPUs. They include for graph analytics, [27], [42], [58], [61], [63] for graph mining

[10], [12], [21], [31], [43], [44], [60], [66], [68], [71]. These frameworks all suffer from poor memory hierarchy performance due to block-based data movement and limited reuse that produce high cache miss rates and low memory parallelism.

Software frameworks have also been designed for GPUs [13], [26], [69]. They suffer from poor compute performance (SIMT divergence and varied memory latency), block-based data movement and limited reuse. GPUs also suffer from limited memory capacity. UpDown enables data movement control (size, intelligence, parallelism), flexible MIMD parallelism (relative to GPUs) combined with high hardware parallelism (vs CPUs). Together, these enable robust scalable performance across diverse graph computing workloads.

## VIII. SUMMARY AND FUTURE WORK

We have presented UpDown, a programmable accelerator, and demonstrated that it achieves scalable, high performance across a breadth of graph computations. Specifically, UpDown outperforms CPU-based software systems by 116x, and prior accelerators by 4x or more. While we only studied a diverse set of 6 applications, the evidence of 197 software-customized events suggests broad applicability. UpDown achieves performance with a set of novel architecture mechanisms that have general applicability, beyond graph applications.

Promising directions for future work include: 1) exploring use of UpDown beyond graph computations; 2) exploring greater interaction with the CPU for applications with data reuse; 3) consideration of different memory system designs.

## ACKNOWLEDGMENTS

This research is based upon work supported by the Office of the Director of National Intelligence (ODNI), Intelligence Advanced Research Projects Activity (IARPA), through the Advanced Graphical Intelligence Logical Computing Environment (AGILE) research program, under Army Research Office (ARO) contract number W911NF22C0082. The views and conclusions contained herein are those of the authors and should not be interpreted as necessarily representing the official policies or endorsements, either expressed or implied, of the ODNI, IARPA, or the U.S. Government.

This work is also supported in part by NSF Grant CNS-1907863 and a Computation Innovation Fellows Award.

## REFERENCES

[1] "Synopsys teaching resources," https://www.synopsys.com/community/university-program/teaching-resources.html.

[2] J. Ahn, S. Hong, S. Yoo, O. Mutlu, and K. Choi, "A scalable processing-in-memory accelerator for parallel graph processing," in *Proceedings of the 42nd Annual International Symposium on Computer Architecture*. Portland Oregon: ACM, Jun. 2015, pp. 105–117.

[3] T. Allen and R. Ge, "In-depth analyses of unified virtual memory system for gpu accelerated computing," in *Proceedings of the International Conference for High Performance Computing, Networking, Storage and Analysis*, ser. SC '21. New York, NY, USA: Association for Computing Machinery, 2021.

[4] T. E. Anderson, B. N. Bershad, E. D. Lazowska, and H. M. Levy, "Scheduler activations: Effective kernel support for the user-level management of parallelism," vol. 25, no. 5, p. 95–109, sep 1991.

[5] S. Beamer, K. Asanovic, and D. Patterson, "Locality exists in graph processing: Workload characterization on an ivy bridge server," in *2015 IEEE International Symposium on Workload Characterization*, 2015, pp. 56–65.

[6] M. Besta, R. Kanakagiri, G. Kwasniewski, R. Ausavarungnirun, J. Beránek, K. Kanellopoulos, K. Janda, Z. Vonarburg-Shmaria, L. Gianinazzi, I. Stefan, J. G. Luna, J. Golinowski, M. Copik, L. Kapp-Schwoerer, S. Di Girolamo, N. Blach, M. Konieczny, O. Mutlu, and T. Hoefler, "SISA: Set-centric instruction set architecture for graph mining on processing-in-memory systems," in *MICRO-54: 54th Annual IEEE/ACM International Symposium on Microarchitecture*, ser. MICRO '21. Association for Computing Machinery, pp. 282–297. [Online]. Available: https://doi.org/10.1145/3466752.3480133

[7] N. Binkert, B. Beckmann, G. Black, S. K. Reinhardt, A. Saidi, A. Basu, J. Hestness, D. R. Hower, T. Krishna, S. Sardashti, R. Sen, K. Sewell, M. Shoaib, N. Vaish, M. D. Hill, and D. A. Wood, "The gem5 simulator," *ACM SIGARCH Computer Architecture News*, vol. 39, no. 2, pp. 1–7, May 2011.

[8] S. Borkar, R. Cohn, G. Cox, T. Gross, H. T. Kung, M. Lam, M. Levine, B. Moore, W. Moore, C. Peterson, J. Susman, J. Sutton, J. Urbanski, and J. Webb, "Supporting systolic and memory communication in iwarp," in *Proceedings of the 17th Annual International Symposium on Computer Architecture*, ser. ISCA '90. New York, NY, USA: Association for Computing Machinery, 1990, p. 70–81. [Online]. Available: https://doi.org/10.1145/325164.325116

[9] D. Brooks, "What's the future of technology scaling?" 2018. [Online]. Available: https://www.sigarch.org/whats-the-future-of-technology-scaling/

[10] H. Chen, M. Liu, Y. Zhao, X. Yan, D. Yan, and J. Cheng, "G-miner: An efficient task-oriented graph mining system," in *Proceedings of the Thirteenth EuroSys Conference*, ser. EuroSys '18. New York, NY, USA: Association for Computing Machinery, 2018. [Online]. Available: https://doi.org/10.1145/3190508.3190545

[11] Q. Chen, B. Tian, and M. Gao, "FINGERS: exploiting fine-grained parallelism in graph mining accelerators," in *Proceedings of the 27th ACM International Conference on Architectural Support for Programming Languages and Operating Systems*. ACM, pp. 43–55. [Online]. Available: https://dl.acm.org/doi/10.1145/3503222.3507730

[12] X. Chen, R. Dathathri, G. Gill, L. Hoang, and K. Pingali, "Sandslash: a two-level framework for efficient graph pattern mining," in *Proceedings of the ACM International Conference on Supercomputing*, ser. ICS '21. Association for Computing Machinery, pp. 378–391. [Online]. Available: https://dl.acm.org/doi/10.1145/3447818.3460359

[13] X. Chen, R. Dathathri, G. Gill, and K. Pingali, "Pangolin: an efficient and flexible graph mining system on CPU and GPU," vol. 13, no. 8, pp. 1190–1205. [Online]. Available: https://dl.acm.org/doi/10.14778/3389133.3389137

[14] X. Chen, T. Huang, S. Xu, T. Bourgeat, C. Chung, and A. Arvind, "FlexMiner: A pattern-aware accelerator for graph pattern mining," in *2021 ACM/IEEE 48th Annual International Symposium on Computer Architecture (ISCA)*. IEEE, pp. 581–594. [Online]. Available: https://ieeexplore.ieee.org/document/9499844/

[15] A. Chien, A. Rajasukumar, M. Nourian, Y. Wang, T. Su, C. Zou, and Y. Fang, "UpDown Accelerator Instruction Set Architecture (ISA) v2.4," University of Chicago, Technical Report TR-2024-03, July 2024. [Online]. Available: https://newtraell.cs.uchicago.edu/research/publications/techreports/TR-2024-03

[16] Community, "The graph 500:large scale graph processing benchmarks." [Online]. Available: https://graph500.org/

[17] I. Cuttress, "The Intel Skylake-X Review: Core i9 7900x, i7 7820x and i7 7800x," 2017. [Online]. Available: https://www.anandtech.com/show/11550/the-intel-skylake-review-core-i9-7900x-i7-7820x-and-i7-7800x-tested

[18] V. Dadu, S. Liu, and T. Nowatzki, "Polygraph: Exposing the value of flexibility for graph processing accelerators," in *2021 ACM/IEEE 48th Annual International Symposium on Computer Architecture (ISCA)*, 2021, pp. 595–608.

[19] W. J. Dally, L. Chao, A. Chien, S. Hassoun, W. Horwat, J. Kaplan, P. Song, B. Totty, and S. Wills, "Architecture of a message-driven processor," in *Proceedings of the 14th Annual International Symposium on Computer Architecture*, ser. ISCA '87. New York, NY, USA: Association for Computing Machinery, 1987, p. 189–196.

[20] T. A. Davis, "Algorithm 1000: Suitesparse:graphblas: Graph algorithms in the language of sparse linear algebra," *ACM Trans. Math. Softw.*, vol. 45, no. 4, dec 2019. [Online]. Available: https://doi.org/10.1145/3322125

[21] V. Dias, C. H. C. Teixeira, D. Guedes, W. Meira, and S. Parthasarathy, "Fractal: A general-purpose graph pattern mining system," in *Proceedings of the 2019 International Conference on Management of Data*, ser. SIGMOD '19. Association for Computing Machinery, pp. 1357–1374. [Online]. Available: https://dl.acm.org/doi/10.1145/3299869.3319875

[22] M. Elseidy, E. Abdelhamid, S. Skiadopoulos, and P. Kalnis, "Grami: Frequent subgraph and pattern mining in a single large graph," *Proceedings of the VLDB Endowment*, vol. 7, no. 7, pp. 517–528, 2014.

[23] P. ERDdS and A. R&wi, "On random graphs i," *Publ. math. debrecen*, vol. 6, no. 290-297, p. 18, 1959.

[24] Y. Fang, C. Zou, A. J. Elmore, and A. A. Chien, "UDP: a programmable accelerator for extract-transform-load workloads and more," in *Proceedings of the 50th Annual IEEE/ACM International Symposium on Microarchitecture*, ser. MICRO-50 '17. New York, NY, USA: Association for Computing Machinery, Oct. 2017, pp. 55–68.

[25] J.-P. Fricker, "The cerebras cs-2: Designing an ai accelerator around the world's largest 2.6 trillion transistor chip," in *Proceedings of the 2022 International Symposium on Physical Design*, ser. ISPD '22. New York, NY, USA: Association for Computing Machinery, 2022, p. 71.

[26] A. Gharaibeh, L. Beltrão Costa, E. Santos-Neto, and M. Ripeanu, "A yoke of oxen and a thousand chickens for heavy lifting graph processing," in *Proceedings of the 21st International Conference on Parallel Architectures and Compilation Techniques*, ser. PACT '12. New York, NY, USA: Association for Computing Machinery, 2012, p. 345–354. [Online]. Available: https://doi.org/10.1145/2370816.2370866

[27] J. E. Gonzalez, Y. Low, H. Gu, D. Bickson, and C. Guestrin, "PowerGraph: Distributed Graph-Parallel computation on natural graphs," in *10th USENIX Symposium on Operating Systems Design and Implementation (OSDI 12)*. Hollywood, CA: USENIX Association, Oct. 2012, pp. 17–30.

[28] C.-Y. Gui, L. Zheng, B. He, C. Liu, X.-Y. Chen, X.-F. Liao, and H. Jin, "A Survey on Graph Processing Accelerators: Challenges and Opportunities," *Journal of Computer Science and Technology*, vol. 34, no. 2, pp. 339–371, Mar. 2019.

[29] T. J. Ham, L. Wu, N. Sundaram, N. Satish, and M. Martonosi, "Graphicionado: a high-performance and energy-efficient accelerator for graph analytics," in *The 49th Annual IEEE/ACM International Symposium on Microarchitecture*, ser. MICRO-49. Taipei, Taiwan: IEEE Press, Oct. 2016, pp. 1–13.

[30] T. Hočevar and J. Demšar, "A combinatorial approach to graphlet counting," *Bioinformatics*, vol. 30, no. 4, pp. 559–565, 12 2014. [Online]. Available: https://doi.org/10.1093/bioinformatics/btt717

[31] K. Jamshidi, R. Mahadasa, and K. Vora, "Peregrine: A pattern-aware graph mining system," in *Proceedings of the Fifteenth European Conference on Computer Systems*, pp. 1–16. [Online]. Available: http://arxiv.org/abs/2004.02369

[32] O. Kalinsky, B. Kimelfeld, and Y. Etsion, "The TrieJax architecture: Accelerating graph operations through relational joins." [Online]. Available: http://arxiv.org/abs/1905.08021

[33] P. M. Kogge, "Jaccard coefficients as a potential graph benchmark," in *2016 IEEE International Parallel and Distributed Processing Symposium Workshops (IPDPSW)*, 2016, pp. 921–928.

[34] H. Kwak, C. Lee, H. Park, and S. Moon, "What is twitter, a social network or a news media?" in *Proceedings of the 19th International Conference on World Wide Web*, ser. WWW '10. New York, NY, USA: Association for Computing Machinery, 2010, p. 591–600. [Online]. Available: https://doi.org/10.1145/1772690.1772751

[35] A. Kyrola, G. Blelloch, and C. Guestrin, "GraphChi: Large-Scale graph computation on just a PC," in *10th USENIX Symposium on Operating Systems Design and Implementation (OSDI 12)*. Hollywood, CA: USENIX Association, Oct. 2012, pp. 31–46.

[36] C. E. Leiserson, N. C. Thompson, J. S. Emer, B. C. Kuszmaul, B. W. Lampson, D. Sanchez, and T. B. Schardl, "There's plenty of room at the top: What will drive computer performance after moore's law?" *Science*, vol. 368, no. 6495, p. eaam9744, 2020. [Online]. Available: https://www.science.org/doi/abs/10.1126/science.aam9744

[37] J. Leskovec, J. Kleinberg, and C. Faloutsos, "Graphs over time: densification laws, shrinking diameters and possible explanations," in *Proceedings of the eleventh ACM SIGKDD international conference on Knowledge discovery in data mining*, 2005, pp. 177–187.

[38] J. Leskovec, K. J. Lang, A. Dasgupta, and M. W. Mahoney, "Community structure in large networks: Natural cluster sizes and the absence of large well-defined clusters," *Internet Mathematics*, vol. 6, no. 1, pp. 29–123, 2009.

[39] S. Li, Z. Yang, D. Reddy, A. Srivastava, and B. Jacob, "DRAMsim3: A Cycle-Accurate, Thermal-Capable DRAM Simulator," *IEEE Computer Architecture Letters*, vol. 19, no. 2, pp. 106–109, Jul. 2020, conference Name: IEEE Computer Architecture Letters.

[40] G. Linden, B. Smith, and J. York, "Amazon.com recommendations: item-to-item collaborative filtering," *IEEE Internet Computing*, vol. 7, no. 1, pp. 76–80, 2003.

[41] J. Lowe-Power, A. M. Ahmad, A. Akram, M. Alian, R. Amslinger, M. Andreozzi, A. Armejach, N. Asmussen, B. Beckmann, S. Bharadwaj, G. Black, G. Bloom, B. R. Bruce, D. R. Carvalho, J. Castrillon, L. Chen, N. Derumigny, S. Diestelhorst, W. Elsasser, C. Escuin, M. Fariborz, A. Farmahini-Farahani, P. Fotouhi, R. Gambord, J. Gandhi, D. Gope, T. Grass, A. Gutierrez, B. Hanindhito, A. Hansson, S. Haria, A. Harris, T. Hayes, A. Herrera, M. Horsnell, S. A. R. Jafri, R. Jagtap, H. Jang, R. Jeyapaul, T. M. Jones, M. Jung, S. Kannoth, H. Khaleghzadeh, Y. Kodama, T. Krishna, T. Marinelli, C. Menard, A. Mondelli, M. Moreto, T. Mück, O. Naji, K. Nathella, H. Nguyen, N. Nikoleris, L. E. Olson, M. Orr, B. Pham, P. Prieto, T. Reddy, A. Roelke, M. Samani, A. Sandberg, J. Setoain, B. Shingarov, M. D. Sinclair, T. Ta, R. Thakur, G. Travaglini, M. Upton, N. Vaish, I. Vougioukas, W. Wang, Z. Wang, N. Wehn, C. Weis, D. A. Wood, H. Yoon, and E. F. Zulian, "The gem5 Simulator: Version 20.0+," Sep. 2020, arXiv:2007.03152 [cs].

[42] G. Malewicz, M. H. Austern, A. J. Bik, J. C. Dehnert, I. Horn, N. Leiser, and G. Czajkowski, "Pregel: A system for large-scale graph processing," in *Proceedings of the 2010 ACM SIGMOD International Conference on Management of Data*, ser. SIGMOD '10. New York, NY, USA: Association for Computing Machinery, 2010, p. 135–146.

[43] D. Mawhirter, S. Reinehr, C. Holmes, T. Liu, and B. Wu, "GraphZero: A high-performance subgraph matching system," vol. 55, no. 1, pp. 21–37. [Online]. Available: https://dl.acm.org/doi/10.1145/3469379.3469383

[44] D. Mawhirter and B. Wu, "AutoMine: harmonizing high-level abstraction and high performance for graph mining," in *Proceedings of the 27th ACM Symposium on Operating Systems Principles*. ACM, pp. 509–523.

[45] J. D. McCalpin, "Bandwidth limits in the intel xeon max (sapphire rapids with hbm) processors," in *High Performance Computing: ISC High Performance 2023 International Workshops, Hamburg, Germany, May 21–25, 2023, Revised Selected Papers*. Berlin, Heidelberg: Springer-Verlag, 2023, p. 403–413. [Online]. Available: https://doi.org/10.1007/978-3-031-40843-4_30

[46] R. R. McCune, T. Weninger, and G. Madey, "Thinking Like a Vertex: A Survey of Vertex-Centric Frameworks for Large-Scale Distributed Graph Processing," *ACM Computing Surveys*, vol. 48, no. 2, pp. 1–39, Nov. 2015.

[47] H. Mujtaba, "Intel sapphire rapids '4th gen xeon' cpu delidded by der8auer, unveils extreme core count die with 56 golden cove cores," *WCCFTech*, 2022. [Online]. Available: https://wccf.tech/189cd

[48] N. Muralimanohar, R. Balasubramonian, and N. P. Jouppi, "Cacti 6.0: A tool to model large caches," *HP laboratories*, vol. 27, p. 28, 2009.

[49] O. Mutlu, S. Ghose, J. Gómez-Luna, and R. Ausavarungnirun, "A modern primer on processing in memory," in *Emerging Computing: From Devices to Systems: Looking Beyond Moore and Von Neumann*, M. M. S. Aly and A. Chattopadhyay, Eds. Springer Nature Singapore, pp. 171–243. [Online]. Available: https://doi.org/10.1007/978-981-16-7487-7_7

[50] L. Nai, R. Hadidi, J. Sim, H. Kim, P. Kumar, and H. Kim, "GraphPIM: Enabling Instruction-Level PIM Offloading in Graph Computing Frameworks," in *2017 IEEE International Symposium on High Performance Computer Architecture (HPCA)*, Feb. 2017, pp. 457–468, iSSN: 2378-203X.

[51] M. E. J. Newman, "The structure and function of complex networks," *SIAM Review*, vol. 45, no. 2, p. 167–256, Jan. 2003. [Online]. Available: http://dx.doi.org/10.1137/S003614450342480

[52] R. S. Nikhil, G. M. Papadopoulos, and Arvind, "T: A multithreaded massively parallel architecture," in *Proceedings of the 19th Annual International Symposium on Computer Architecture*, ser. ISCA '92. New York, NY, USA: Association for Computing Machinery, 1992, p. 156–167.

[53] M. Orenes-Vera, E. Tureci, D. Wentzlaff, and M. Martonosi, "Dalorex: A data-local program execution and architecture for memory-bound applications," in *2023 IEEE International Symposium on High-Performance Computer Architecture (HPCA)*, 2023, pp. 718–730.

[54] L. Page, S. Brin, R. Motwani, and T. Winograd, "The pagerank citation ranking: Bringing order to the web." Stanford InfoLab, Technical Report 1999-66, November 1999. [Online]. Available: http://ilpubs.stanford.edu:8090/422/

[55] S. Rahman, N. Abu-Ghazaleh, and R. Gupta, "GraphPulse: An Event-Driven Hardware Accelerator for Asynchronous Graph Processing," in *2020 53rd Annual IEEE/ACM International Symposium on Microarchitecture (MICRO)*, Oct. 2020, pp. 908–921.

[56] T. Reddy, K. Murugavel, and J. Lowe-Power, "gem5 skylake config," 2020. [Online]. Available: https://github.com/darchr/gem5-skylake-config/

[57] R. A. Rossi and N. K. Ahmed, "The network data repository with interactive graph analytics and visualization," in *AAAI*, 2015. [Online]. Available: https://networkrepository.com

[58] A. Roy, I. Mihailovic, and W. Zwaenepoel, "X-stream: Edge-centric graph processing using streaming partitions," in *Proceedings of the Twenty-Fourth ACM Symposium on Operating Systems Principles*, ser. SOSP '13. New York, NY, USA: Association for Computing Machinery, 2013, p. 472–488. [Online]. Available: https://doi.org/10.1145/2517349.2522740

[59] B. C. Schwedock, P. Yoovidhya, J. Seibert, and N. Beckmann, "täkō: a polymorphic cache hierarchy for general-purpose optimization of data movement," in *Proceedings of the 49th Annual International Symposium on Computer Architecture*. New York New York: ACM, Jun. 2022, pp. 42–58.

[60] T. Shi, M. Zhai, Y. Xu, and J. Zhai, "GraphPi: high performance graph pattern matching through effective redundancy elimination," in *Proceedings of the International Conference for High Performance Computing, Networking, Storage and Analysis*, ser. SC '20. IEEE Press, pp. 1–14.

[61] J. Shun and G. E. Blelloch, "Ligra: A Lightweight Graph Processing Framework for Shared Memory," in *ACM SIGPLAN Symposium on Principles Practice of Parallel Programming*, 2013, pp. 135–146.

[62] A. Stillmaker and B. Baas, "Scaling equations for the accurate prediction of cmos device performance from 180nm to 7nm," *Integration*, vol. 58, pp. 74–81, 2017.

[63] N. Sundaram, N. Satish, M. M. A. Patwary, S. R. Dulloor, M. J. Anderson, S. G. Vadlamudi, D. Das, and P. Dubey, "GraphMat: high performance graph analytics made productive," *Proceedings of the VLDB Endowment*, vol. 8, no. 11, pp. 1214–1225, Jul. 2015.

[64] N. Talati, H. Ye, Y. Yang, L. Belayneh, K.-Y. Chen, D. Blaauw, T. Mudge, and R. Dreslinski, "NDMiner: accelerating graph pattern mining using near data processing," in *Proceedings of the 49th Annual International Symposium on Computer Architecture*. ACM, pp. 146–159. [Online]. Available: https://dl.acm.org/doi/10.1145/3470496.3527437

[65] M. B. Taylor, W. Lee, J. Miller, D. Wentzlaff, I. Bratt, B. Greenwald, H. Hoffmann, P. Johnson, J. Kim, J. Psota, A. Saraf, N. Shnidman, V. Strumpen, M. Frank, S. Amarasinghe, and A. Agarwal, "Evaluation of the raw microprocessor: An exposed-wire-delay architecture for ilp and streams," in *Proceedings of the 31st Annual International Symposium on Computer Architecture*, ser. ISCA '04. USA: IEEE Computer Society, 2004, p. 2.

[66] C. H. C. Teixeira, A. J. Fonseca, M. Serafini, G. Siganos, M. J. Zaki, and A. Aboulnaga, "Arabesque: a system for distributed graph mining," in *Proceedings of the 25th Symposium on Operating Systems Principles*, ser. SOSP '15. Association for Computing Machinery, pp. 425–440. [Online]. Available: https://dl.acm.org/doi/10.1145/2815400.2815410

[67] A. Tumeo and J. Feo, "Irregular applications: From architectures to algorithms [guest editors; introduction]," *Computer*, vol. 48, no. 08, pp. 14–16, aug 2015.

[68] K. Wang, Z. Zuo, J. Thorpe, T. Q. Nguyen, and G. H. Xu, "{RStream}: Marrying relational algebra with streaming for efficient graph mining on a single machine," pp. 763–782. [Online]. Available: https://www.usenix.org/conference/osdi18/presentation/wang

[69] Y. Wang, A. Davidson, Y. Pan, Y. Wu, A. Riffel, and J. D. Owens, "Gunrock: a high-performance graph processing library on the gpu," *SIGPLAN Not.*, vol. 51, no. 8, feb 2016. [Online]. Available: https://doi.org/10.1145/3016078.2851145

[70] Y. Wu, J. Zhu, W. Wei, L. Chen, L. Wang, S. Wei, and L. Liu, "Shogun: A task scheduling framework for graph mining accelerators," in *Proceedings of the 50th Annual International Symposium on Computer Architecture*. ACM, pp. 1–15. [Online]. Available: https://dl.acm.org/doi/10.1145/3579371.3589086

[71] D. Yan, G. Guo, M. M. Rahman Chowdhury, M. Tamer Özsu, W.-S. Ku, and J. C. S. Lui, "G-thinker: A distributed framework for mining subgraphs in a big graph," in *2020 IEEE 36th International Conference on Data Engineering (ICDE)*, pp. 1369–1380, ISSN: 2375-026X.

[72] J. Yang and J. Leskovec, "Defining and evaluating network communities based on ground-truth," in *Proceedings of the ACM SIGKDD Workshop on Mining Data Semantics*, 2012, pp. 1–8.

[73] P. Yao, L. Zheng, Z. Zeng, Y. Huang, C. Gui, X. Liao, H. Jin, and J. Xue, "A locality-aware energy-efficient accelerator for graph mining applications," in *2020 53rd Annual IEEE/ACM International Symposium on Microarchitecture (MICRO)*, pp. 895–907.

[74] X. Yu, C. J. Hughes, N. Satish, and S. Devadas, "IMP: indirect memory prefetcher," in *Proceedings of the 48th International Symposium on Microarchitecture*. Waikiki Hawaii: ACM, Dec. 2015, pp. 178–190.

[75] D. Zhang, X. Ma, M. Thomson, and D. Chiou, "Minnow: Lightweight Offload Engines for Worklist Management and Worklist-Directed Prefetching," *ACM SIGPLAN Notices*, vol. 53, no. 2, pp. 593–607, Mar. 2018.

[76] M. Zhang, Y. Zhuo, C. Wang, M. Gao, Y. Wu, K. Chen, C. Kozyrakis, and X. Qian, "GraphP: Reducing Communication for PIM-Based Graph Processing with Efficient Data Partition," in *2018 IEEE International Symposium on High Performance Computer Architecture (HPCA)*, Feb. 2018, pp. 544–557, iSSN: 2378-203X.

[77] Y. Zhang, X. Liao, H. Jin, L. He, B. He, H. Liu, and L. Gu, "DepGraph: A Dependency-Driven Accelerator for Efficient Iterative Graph Processing," in *2021 IEEE International Symposium on High-Performance Computer Architecture (HPCA)*, Feb. 2021, pp. 371–384, iSSN: 2378-203X.

[78] Y. Zhuo, C. Wang, M. Zhang, R. Wang, D. Niu, Y. Wang, and X. Qian, "GraphQ: Scalable PIM-Based Graph Processing," in *Proceedings of the 52nd Annual IEEE/ACM International Symposium on Microarchitecture*, ser. MICRO '52. New York, NY, USA: Association for Computing Machinery, Oct. 2019, pp. 712–725.